\documentclass{sig-alternate}

\usepackage[ruled]{algorithm2e}
\usepackage{textcomp}
\usepackage{amsmath, amsfonts,amssymb}
\usepackage{subfigure}
\usepackage{wasysym} %other smileys

\SetAlFnt{\small}
\SetAlCapFnt{\small}
\SetAlCapNameFnt{\small}
\SetAlCapHSkip{0pt}
\IncMargin{-\parindent}

\newenvironment{packed_item}{
\begin{itemize}
  \setlength{\itemsep}{1pt}
  \setlength{\parskip}{0pt}
  \setlength{\parsep}{0pt}
}{\end{itemize}}

\begin{document}

% \conferenceinfo{CIKM'13,} {Oct. 27 -- Nov. 1, 2013, San Francisco, CA, USA.} 
% \CopyrightYear{2013} 
% \crdata{xxx-x-xxxx-xxxx-xx/xx/xx} 
% \clubpenalty=10000 
% \widowpenalty = 10000
% --- Author Metadata here ---
% \conferenceinfo{SIGIR}{'12 Portland, Oregon, USA}
%\CopyrightYear{2007} % Allows default copyright year (20XX) to be over-ridden - IF NEED BE.
%\crdata{0-12345-67-8/90/01}  % Allows default copyright data (0-89791-88-6/97/05) to be over-ridden - IF NEED BE.
% --- End of Author Metadata ---

\title{GAPfm: Optimal Top-N Recommendations for Graded Relevance Domains
\titlenote{A short version of this manuscript is published at CIKM 2013.}}

\numberofauthors{5} %  in this sample file, there are a *total*
\author{
\alignauthor
Yue Shi\\ \affaddr{Delft University of
Technology}\\
       \email{y.shi@tudelft.nl}
% 2nd. author
\alignauthor
Alexandros Karatzoglou\\
       \affaddr{Telefonica Research}\\
       \email{alexk@tid.es}
% 3rd. author
\alignauthor Linas Baltrunas\\
       \affaddr{Telefonica Research}\\
       \email{linas@tid.es}
\and  % use '\and' if you need 'another row' of author names
% 4th. author
\alignauthor Martha Larson\\
       \affaddr{Delft University of Technology}\\
       \email{m.a.larson@tudelft.nl}
\alignauthor Alan Hanjalic\\
       \affaddr{Delft University of Technology}\\
       \email{a.hanjalic@tudelft.nl}
% \alignauthor Nuria Oliver\\
%        \affaddr{Telefonica Research}\\
%        \email{nuriao@tid.es}
}

% \numberofauthors{1} %  in this sample file, there are a *total*
% \author{
% \alignauthor
% Yue Shi$^a$, Alexandros Karatzoglou$^b$, Linas
% Baltrunas$^b$, Martha Larson$^a$, \\Alan Hanjalic$^a$\\
% \affaddr{$^a$Delft University of Technology, Netherlands; $^b$Telefonica
% Research, Spain}\\ \email{$^a$\{y.shi, m.a.larson, a.hanjalic\}@tudelft.nl,
% $^b$\{alexk, linas\}@tid.es}
% }
% 2nd. author
% \alignauthor
% \\
%        \affaddr{}\\
%        \email{alexk@tid.es}
% % 3rd. author
% \alignauthor \\
%        \affaddr{Telefonica Research}\\
%        \email{linas@tid.es}
% \and  % use '\and' if you need 'another row' of author names
% % 4th. author
% \alignauthor \\
%        \affaddr{Delft University of Technology}\\
%        \email{m.a.larson@tudelft.nl}
% \alignauthor \\
%        \affaddr{Delft University of Technology}\\
%        \email{a.hanjalic@tudelft.nl}
% \alignauthor \\
%        \affaddr{Telefonica Research}\\
%        \email{nuriao@tid.es}

\maketitle
\begin{abstract}
Recommender systems are frequently used in domains in which users express their
preferences in the form of graded judgments, such as ratings. If accurate top-N
recommendation lists are to be produced for such graded relevance domains, it is
critical to generate a ranked list of recommended items directly rather than
predicting ratings. Current techniques choose one of two sub-optimal approaches:
either they optimize for a binary metric such as Average Precision, which
discards information on relevance grades, or they optimize for Normalized
Discounted Cumulative Gain (NDCG), which ignores the dependence of an item's
contribution on the relevance of more highly ranked items. 

In this paper, we
address the shortcomings of existing approaches by proposing the Graded Average
Precision factor model ({\it GAPfm}), a latent factor model that is particularly
suited to the problem of top-N recommendation in domains with graded relevance
data. The model optimizes for Graded Average Precision, a metric that has been
proposed recently for assessing the quality of ranked results list for graded
relevance. {\it GAPfm} learns a latent factor model by directly optimizing a smoothed
approximation of GAP. {\it GAPfm}'s advantages are twofold: it maintains full
information about graded relevance and also addresses the limitations of models
that optimize NDCG. Experimental results show that {\it GAPfm} achieves substantial
improvements on the top-N recommendation task, compared to several
state-of-the-art approaches. In order to ensure that {\it GAPfm} is able to scale to
very large data sets, we propose a fast learning algorithm that uses an adaptive
item selection strategy. A final experiment shows that {\it GAPfm} is useful not only
for generating recommendation lists, but also for ranking a given list of rated
items.
\end{abstract}
% \vspace{-3mm}
{\small
% \category{H.3.3}{Information Storage and Retrieval}{Information Search
% and Retrieval}[Information Filtering]
% \vspace{-2mm}
}

\section{Introduction}
\label{sec:intro}
%Text before ML changes:
%Recommendation technology has been widely adopted by many online
%services in recent years, to alleviate users from the massive information overload~\cite{Adomavicius2005}. 
%Collaborative filtering (CF) is one of the most popular and successful techniques for recommender 
%systems~\cite{Ekstrand2011} and is deployed, for example, by Amazon and Netflix. The core idea behind 
%CF is that users whose past interests were similar will also share common interests in future. 
%The user's interest is inferred by the user interaction patterns with
%the items either explicitly or implicitly.
%In explicit feedback, users are asked/allowed 
%to explicitly rate the items that have been
%purchased/consumed, using a pre-defined Likert scale (graded relevance),
%e.g., 1-5 stars in Netflix movie recommendation site. Higher grade
%indicates higher preference/relevance of the item.
%In implicit feedback~\cite{Hu2008}, users interact with items by
%downloading, purchasing, and their preferences are deduced from the
%interaction patterns. In this paper we propose a new CF model for the
%case of explicit/graded relevance data. 

Recommendation technology has been widely adopted by many online
services in recent years, to help relieve users of massive information overload~\cite{Adomavicius2005}. 
Collaborative filtering (CF) is one of the most popular and successful techniques for recommender 
systems~\cite{Ekstrand2011} and is deployed, for example, by Amazon and Netflix. The core idea behind 
CF is that users whose past interests were similar will also share common interests in future. 
Users' interests are inferred from patterns of interaction, either explicit or implicit, of users with items.
In a typical case involving explicit feedback, users are asked/allowed 
to explicitly rate items, using a pre-defined rating scale (graded relevance),
e.g., 1-5 stars on the Netflix movie recommendation site. A higher grade
indicates stronger preference for the item, reflecting a higher relevance of the item with respect to the user.
In a typical case involving implicit feedback~\cite{Hu2008}, users interact with items by
downloading or purchasing, and their preferences are deduced from the
resulting patterns. In this paper we propose a new CF model for the
case of graded relevance data. 

%Text before ML changes:
%One way to measure the fit of a learned model for graded relevance
%data (e.g. ratings) is to use
%a metric such as Root-Mean-Square Error. This metric was adopted as
%the evaluation metric in the Netflix Prize
%contest\footnote{http://www.netflixprize.com/}. However, it is now  widely 
%recognized that the recommendation approaches optimized for minimizing the error
%rate usually achieve poor performance for the top-N recommendation
%task~\cite{Cremonesi2010,Gunawardana2009}. 
%Moreover, users generally pay attention only to a short list of $N$ recommended items.
%It is thus much more useful to focus the modeling power in trying to
%make this short list of \emph{top-N} items as relevant as possible and not to
%accurately predict ratings of non-relevant items. Despite the different nature of the underlying data (graded v.s. binary) the ultimate objective of
%CF models in all cases is the same, i.e., \emph{to generate a top-N
%recommendation list of relevant items to individual users.} This task
%is essentially a ranking task, i.e. ranking items according to their
%relevance to the user. An approach thus often used in CF models is to
%optimize for a ranking metric. 

One way to measure the fit of a learned model for graded relevance
data (e.g., ratings) is to use
a metric such as Root-Mean-Square Error. This metric was adopted as
the evaluation metric in the Netflix Prize
contest\footnote{http://www.netflixprize.com/}. However, it is now  widely 
recognized that recommendation approaches optimized to minimize the error
rate usually achieve poor performance on the top-N recommendation
task~\cite{Cremonesi2010,Gunawardana2009}. 
In practice, users focus their attention on only a small number of recommendations, effectively ignoring all but a short list of $N$ recommended items.
For this reason, it is more useful to focus the recommendation model on 
making this short list of \emph{top-N} items as relevant as possible, rather than on
accurately predicting ratings of non-relevant items. Despite the different nature of the underlying data (graded v.s. binary) the ultimate objective of
CF models in all cases is the same, i.e., \emph{to generate a top-N
recommendation list of relevant items to individual users.} This task
is essentially a ranking task, i.e., ranking items according to their
relevance to the user. Consequently, CF models that optimize for a ranking metric are particularly well suited to address it.

Various models use learning to rank~\cite{Liu2009b} techniques 
to optimize binary relevance ranking metrics.
For example, several CF models~\cite{Rendle2009,Shi2012,Shi2012b} compute near optimal ranked lists with respect to the 
Area Under the Curve (AUC), Average Precision (AP)~\cite{Manning2008} and
Reciprocal Rank~\cite{Voorhees1999} metrics. 
However, metrics that are defined to handle binary relevance data are
not directly suitable for graded relevance data. 
% 4U2CHECK: Can we say this?
%In practice, we would need to first convert the graded relevance data to binary by imposing a thresholding relevance (e.g., setting rating 4 as the threshold for the 1-5 scale so that items rated 4 and 5 are treated as relevant). 
%Binary metrics, and CF methods that optimize for these metrics can be used on
%graded relevance data if it is converted to binary relevance by e.g., imposing a
%threshold (e.g., setting rating 4 as the threshold for the 1-5 scale so that items rated 4 and 5 are
%treated as relevant). 
In order to apply binary metrics, and CF methods that optimize for these metrics, to graded relevance data, it is necessary to convert the data to binary relevance data. 
This conversion is generally accomplished by imposing a threshold (e.g., defining rating levels 1-3 as non-relevant and 4-5 as relevant).
% LB above is not clear yet... its too complicated sentence.
% ML I have further edited above to make it clearer
This process has two major drawbacks: 1) we {\it lose grading information}
% ML changed: within the rated items, e.g., items rated with a 5 are more relevant then items
among the rated items, e.g., items rated with a 5 are more relevant than items
rated with a 4. This information is crucial in building precise models.
2) the choice of
%ML changed: the {\it thresholding relevance is arbitrary} and will
the {\it threshold relevance is arbitrary} and will
have an impact on the performance of different recommendation
approaches. 
%thus evaluation of these approaches becomes very difficult. 
% 4U2CHECK: This argumentation is important, so I tried to re-introduce it.
%For this reason, advances in the area of top-N recommendation that use CF
%optimized for binary relevance metrics, cannot be directly transferred to
%recommendation domains involving graded relevance data.

%Another well known alternative metric in the area of IR is the Normalized
% 4U2CHECK: "alternative" sounds strange if you already write "another"
A well-known metric in the area of information retrieval (IR) is Normalized
Discounted Cumulative Gain (NDCG)~\cite{Jarvelin2002}, which can be used to
measure the performance of ranked results with graded relevance and is often
used for evaluating recommender
systems~\cite{Balakrishnan2012,Liu2008,Liu2009,Volkovs2012,Weimer2007}. NDCG is
dependent on both the grades and the positions of the items in the ranked list.
However, NDCG is a so-called ``non-cascade" metric. Under ``cascade metrics",
such as Average precision and Reciprocal Rank, the contribution of a given item has a
dependence on the relevance of higher ranked items. Instead, NDCG assumes
independence between the items in the ranked list, i.e., each item contributes
to the quality of the ranked list solely based on its own grade and position,
while ignoring the impact of items that are ranked above it. The
``non-cascade" nature of NDCG, has recently drawn criticism from authors who
point out the advantages of cascade metrics~\cite{Chapelle2009,Robertson2010}.

Graded Average Precision (GAP)~\cite{Robertson2010} has been proposed as
a generalized version of Average Precision in the use scenario with graded
relevance data. GAP, being similar to Average Precision, reflects the
overall quality of the top-N ranked items. 
%ML changed: Moreover, it inherits all the desirable properties that AP has:
%top-heavy bias, highly informative, nice probabilistic interpretation,
%and an underlying theoretical basis~\cite{Robertson2010}.
Moreover, it inherits all the desirable properties of AP:
top-heavy bias, high informativeness, elegant probabilistic interpretation,
and solid underlying theoretical basis~\cite{Robertson2010}.
% 4U2CHECK: 
In this paper, we
%propose a new CF approach, latent factor model for graded average precision (\textbf{{\it GAPfm}}), that learns a latent factor model, i.e., latent factors
propose a new CF approach, i.e., a latent factor model for Graded Average
Precision (\textbf{{\it GAPfm}}), that learns latent factors of users and items
so as to directly optimizes GAP of top-N recommendations.
% 4U2CHECK: I would add this sentence a bit later....under the philosophy of first let the readers enjoy what they *are* getting before mentioning what they aren't. I would put it in the section that presents the details of {\it GAPfm}
The contributions in this paper can be summarized as:
\vspace{-2pt}
\begin{packed_item}
\item We propose a novel CF approach that directly optimizes GAP. We show that
{\it GAPfm} outperforms state-of-the-art methods for various evaluation metrics
including GAP, Precision and NDCG.
\item We conduct a theoretical analysis of the smoothed approximation of GAP and support its validity. 
%\item We conduct a theoretical analysis of the smoothed approximation of GAP and support its validity. 
% 4U2CHECK: do you mean ", which supports its validity" Otherwise it seems like you are doing something additional to support the validity.
\item We provide a learning algorithm that scales linearly with the number of
observed grades in the dataset. Moreover, we further exploit properties of GAP
and provide a sub-linear complexity learning algorithm that is suitable for large
data sets.
% 4U2CHECK: do you mean "to provide" instead of "and provide"?
\end{packed_item}
\vspace{-2pt}
The reminder of the paper is organized as follows: in Section~\ref{sec:rw} we
discuss previous research contributions that are related to our approach
proposed in this paper. Then, in Section~\ref{sec:prob}, we introduce the notation and terminology
adopted throughout the paper, after which, in
Section~\ref{sec:lfmgap}, we present the details of {\it GAPfm}. The
experimental evaluation is presented in Section~\ref{sec:exp}. 
% 4U2CHECK: I changed the wording of the following sentence a bit:
Finally,
Section~\ref{sec:conclude} summarizes our contributions and discusses
future work.
%% old below

\section{Related work}
\label{sec:rw}
% 4U2CHECK: I changed the wording of the following sentence a bit
The approach proposed in this paper is rooted in the research areas of collaborative filtering and
learning to rank. In the following, we discuss related contributions in each of
the two areas, and position our work with respect to them.

\textbf{Collaborative Filtering.}
A large portion of recent CF approaches tries to address the rating prediction
problem, as defined in the Netflix Prize contest. CF approaches can
categorized broadly into two categories: memory-based and model-based~\cite{Adomavicius2005}. Memory-based
approaches rely on the similarities between users or items, and generate
% 4U2CHECK: removed "the"
rating predictions by aggregating preference data over similar users
(user-based)~\cite{Herlocker1999} or similar items
(item-based)~\cite{Sarwar2001}. Model-based approaches learn a prediction model based on a set of training data,
and then use the prediction model to generate recommendations for individual
users~\cite{Agarwal2009,Hofmann2004}. Latent factor models (or
more specifically, matrix factorization techniques) have attracted significant
% 4U2CHECK: changed performances -> performance (it doesn't work well with plural)
research attention, due to their superior performance on the rating prediction
problem, as witnessed during the Netflix Prize contest~\cite{Koren2008,Koren2009a}.
% 4U2CHECK: changed where -> were
The methods developed to attack the Netflix Prize were highly
effective for the rating prediction task, but have turned out to have
%ML changed: relatively poor performance in the top-N recommendation task~\cite{Cremonesi2010}.
relatively poor performance on the top-N recommendation task~\cite{Cremonesi2010}.

%%To address the ranking problem in CF, a few contributions have been proposed, recently. 
%A few contributions have been proposed specifically to address the ranking problem in CF.
%Bayesian personalized ranking (BPR)~\cite{Rendle2009} and
%Collaborative Less-is-More Filtering (CLiMF)~\cite{Shi2012b} directly optimize binary relevance
%measures, i.e., Area Under the Curve (AUC) in BPR and Reciprocal Rank in CLiMF, and seek
%to improve top-N recommendations. In a similar spirit,
%TFMAP~\cite{Shi2012} directly optimizes Average Precision for
%context-aware recommendations. All of these methods use implicit
%feedback data. 
% However, as discussed in Section~\ref{sec:intro}, these
%methods are not optimal for graded relevance datasets, since they are not able
%to fully exploit the information encoded in the grade levels.

%To address the ranking problem in CF, a few contributions have been proposed, recently. 
A few contributions have been proposed specifically to address the ranking problem in CF.
Bayesian personalized ranking (BPR)~\cite{Rendle2009} and
Collaborative Less-is-More Filtering (CLiMF)~\cite{Shi2012b} seek to improve top-N recommendation by directly optimize binary relevance
measures, i.e., Area Under the Curve (AUC) in BPR and Reciprocal Rank in CLiMF.
%, and seek to improve top-N recommendations. 
In a similar spirit,
TFMAP~\cite{Shi2012} directly optimizes Average Precision for
%4YS Please check my change to the sentence below
context-aware recommendations. All of these methods use binary implicit-feedback data. 
 However, as discussed in Section~\ref{sec:intro}, these
methods are not well-suited for graded relevance datasets, since they are not able
to fully exploit the information encoded in the grade levels.

%Original text changed below by ML
%%To address the ranking problem in CF, a few contributions have been proposed, recently. 
%A few contributions have been proposed specifically to address the ranking problem in CF.
%Bayesian personalized ranking (BPR)~\cite{Rendle2009} and
%Collaborative Less-is-More Filtering (CLiMF)~\cite{Shi2012b} directly optimize binary relevance
%measures, i.e., Area Under the Curve (AUC) in BPR and Reciprocal Rank in CLiMF, and seek
%to improve top-N recommendations. In a similar spirit,
%TFMAP~\cite{Shi2012} directly optimizes Average Precision for
%context-aware recommendations. All of these methods use implicit
%feedback data. 
% However, as discussed in Section~\ref{sec:intro}, these
%methods are not optimal for graded relevance datasets, since they are not able
%to fully exploit the information encoded in the grade levels.

Research that deals with the ranking problem for cases involving graded
relevance data includes 
EigenRank~\cite{Liu2008} and probabilistic latent preference
analysis~\cite{Liu2009}, which exploit pair-wise comparisons
between the rated items. Collaborative competitive filtering~\cite{Yang2011b} has further 
advanced the performance of top-N recommendation by imposing
% 4U2CHECK: here, I would like to have "local competition", rather than "a local competition".
local competition, i.e., constraining items that users have seen but not rated to be less preferred that items both seen and rated.
However, none of these methods are designed to optimize for any specific
ranking/evaluation measure.

To our knowledge, the only existing CF approach,
that directly optimizes a graded evaluation measure is
CofiRank~\cite{Weimer2007}, which minimizes a convex upper bound of
the NDCG loss
through matrix factorization. Some of the latest contributions aim at  
enhancing the performance of CofiRank 
and boosting the NDCG score of the ranking
results~\cite{Balakrishnan2012,Volkovs2012}. These approaches are often referred
to as \emph{collaborative ranking}, and are evaluated by their performance on
%ML changed: ranking graded items (Note that these approaches differ from top-N
%recommendation in that they rank a given list of rated items rather than
%generating a recommendation list). As mentioned in Section~\ref{sec:intro}, even
%4YS Please check my change to the sentences below
ranking graded items. Note that these approaches solve a different problem. Instead of addressing the top-N
recommendation task, they rank a list of rated items that has already been given, i.e., pre-specified. As mentioned in Section~\ref{sec:intro}, even
results that are ranked optimally in terms of NDCG may still yield suboptimal top-N
recommendations. The new approach introduced by this paper, GAPfm, directly
optimizes a recently proposed cascade metric GAP. In our experimental
evaluation, we demonstrate that GAPfm outperforms CoFiRank with respect to a
range of conventional top-N evaluation metrics.

\textbf{Learning to Rank.}
%LB I think here we need to add that learning to rank in CF is a new
%trend. Mention ord-rec'2011 and BPR and why we are different.
%AK also not sure this paragragh makes alot of sense we are essentialy
%re-citeing everything above 
The task of learning to rank is to learn a ranking function that is used to rank
documents for given queries~\cite{Liu2009b}. Inspired by the analogy between
query-document relations in IR and user-item relations in recommender systems,
many CF methods were proposed recently 
~\cite{Balakrishnan2012,Hong2012a,Rendle2009,Shi2012,Shi2012b,Volkovs2012,Weimer2007}.
Our work in this paper also falls into this category, and in particular, it is
closely related to one sub-area of learning to rank, i.e., direct
optimization of evaluation metrics.

%AK I think this should be before the CF part 
Note that the key challenge of directly optimizing
evaluation metrics lies in the non-smoothness~\cite{Burges2006} of these measures.
Specifically, these metrics are defined on the rankings and the grades (in the
case of graded relevance data) of documents/items in a list, which are indirectly determined
by the parameters of the ranking function. 
Research contributions have been made to directly optimize evaluation metrics
by exploiting structured estimation techniques~\cite{TsoJoaHofAlt05,Xu2008}
that minimize convex upper bounds of loss functions based on evaluation
measures, e.g., SVM-MAP~\cite{Yue2007} and AdaRank~\cite{Xu2007}.
In addition,
SoftRank~\cite{Taylor2008} and its extensions~\cite{Chapelle2010} were proposed
to use smoothed versions of evaluation measures, which can then be directly
%ML changed: optimized. Our work follows this direction of research, as we propose to optimize a smoothed version of
optimized. Our work can be considered part of this research direction, since our proposed approach optimizes a smoothed version of
%GAP. The difference from previous work is that we integrate GAP optimization
GAP. The difference with previous work is that we integrate GAP optimization
with latent factor models for learning optimal top-N recommendation in
the graded relevance domains. 
% 4U2CHECK: for consistency, I would always refer to it as a "use scenario" not a "use case"
We also contribute a learning algorithm that guarantees that the
proposed approach is highly scalable.
\vspace{-4pt}
\section{Notation and Terminology}
\label{sec:prob}
We denote the graded relevance data from $M$ users to $N$ items as a matrix
$Y^{M\times N}$, in which the entry $y_{mi}$ denotes the grade given by user $m$
to item $i$. Note that we have $y_{mi}\in\{1,2,\ldots,y_{max}\}$, in which
%ML changed: $y_{max}$ is the highest grade. Note that $y_{mi}=0$ indicates that user
$y_{max}$ is the highest grade. Note also that $y_{mi}=0$ indicates that user
$m$'s preference for item $m$ is unknown. $|Y|$ denotes the number of nonzero
entries in $Y$. In addition, $I_{mi}$ serves as an indicator function that is
equal to 1, if $y_{mi}>0$, and 0 otherwise. We use $U^{D\times M}$ to denote the
latent factors of $M$ users, and in particular $U_m$ denotes a $D$-dimensional (column) vector that represents the latent factors for user $m$. Similarly, $V^{D\times N}$ denotes the latent factors of $N$ items and
$V_i$ represents the latent factors of item $i$. Note that the latent factors
in $U$ and $V$ are model parameters that need to be estimated from the data
(i.e., a training set). The relevance between user $m$ and item $i$ is
predicted by the latent factor model, i.e., using the inner product of $U_m$
and $V_i$, as below:
\begin{align}
\label{eqn:tensor}
f_{mi}= \left \langle U_m,V_i\right
\rangle =\sum_{d=1}^{D}U_{md}V_{id}
\end{align}
To produce a ranked list of items for a user $m$ all items are scored
using Eq.~(\ref{eqn:tensor}) and ranked according to the
scores.
In the following, we use $R_{mi}$ to denote the rank position of item $i$ for
user $m$, according to the descending order of predicted relevances of all items
to the user. For example, if the predicted relevance of item $i$ is higher than
that of all the other items for user $m$, i.e., if $f_{mi}>f_{mj}, j=1,2,\ldots,N$ and
$j\neq i$, then $R_{mi}=1$.

Taking into account both the original definition of GAP in~\cite{Robertson2010}
and the notation introduced above, we can rewrite the formulation of GAP for
%ML changed: a ranked item list of user $m$ as follows:
a ranked item list recommended for user $m$ as follows:
\begin{align}
\label{eqn:gap}
GAP_m=& \frac{1}{Z_{m}}\sum_{i=1}^N \frac{I_{mi}}{R_{mi}}
\sum_{j=1}^{N}I_{mj}\mathbb{I}(R_{mj} \leq R_{mi}) \nonumber \\ 
& \Big(
\mathbb{I}(y_{mi}<y_{mj}) \sum_{l=1}^{y_{mi}}\delta_l + \mathbb{I}(y_{mj} \leq y_{mi}) \sum_{l=1}^{y_{mj}}\delta_l \Big)
\end{align}
where $\mathbb{I}(\cdot)$ is an indicator function, which is equal to 1 if the
condition is true, and otherwise 0. $\delta_l$ denotes the thresholding probability~\cite{Robertson2010} that the
user sets as a threshold of relevance at grade $l$, i.e., regarding items with grades equal or larger than $l$ as relevant ones, and items with grades lower than $l$ as irrelevant ones.
$Z_m$ is a constant normalizing coefficient for
user $m$, as defined below:
\begin{align}
\label{eqn:zm}
Z_m=\sum_{l=1}^{y_{max}}n_{ml}\sum_{k=1}^{l}\delta_k
\end{align}
where $n_{ml}$ denotes the number of items rated with grade $l$ by user $m$. 

For notational convenience in the rest of the paper, we substitute the
last term of the parentheses in Eq.~(\ref{eqn:gap}), as shown below:
\begin{align}
\label{eqn:subs}
\beta_{mij}:=
\mathbb{I}(y_{mi}<y_{mj}) \sum_{l=1}^{y_{mi}}\delta_l  + \mathbb{I}(y_{mj} \leq
y_{mi}) \sum_{l=1}^{y_{mj}}\delta_l
\end{align}
We assume that each grade $l$ is an integer ranging from 1 to $y_{max}$, since usually a non-integer grade scale can be
transformed to an integer grade scale by multiplying by a constant factor, e.g.,
the scale of 1 to 5 stars with half star increment can be transformed to the
scale of 1 to 10 stars by multiplying factor 2. 
As suggested in~\cite{Robertson2010}, the value of $\delta_l$ for
each grade needs to be empirically tuned according to the specific use cases. 
% 4U2CHECK: at least change cases -> case, consider writing "scenario"
In this paper, we adopt
an exponential mapping function that maps the grade $l$ to the thresholding
% 4U2CHECK: I tend to write "threshold probability", but this sounds ok, too. 
probability $\delta_l$, as shown in Eq.~(\ref{eqn:gradeprop2}). Note that other
expressions for the definition of $\delta_l$ can be also used, without
influencing the main results on the optimization of GAP, as
presented in the
next section.
\begin{align}
\label{eqn:gradeprop2}
\delta_l=\left\{\begin{matrix}
\frac{2^l-1}{2^{y_{max}}}, & y_{max}>1\\ 
1, & y_{max}=1
\end{matrix}\right.
\end{align}

Using the introduced terminology, the research problem investigated in this
paper can be stated as: \emph{Given a top-N recommendation scenario
  involving graded relevance data $Y$, 
% 4U2CHECK: I would say "Given a top-N recommendation scenario involving graded relevance..."
learn latent factors of users and items, $U$ and $V$, through the
direct optimization of GAP as in
Eq.~(\ref{eqn:gap}) across all the users and items.} 

\section{GAP\lowercase{fm}}
% 4U2CHECK:  I think you need to try to save the lower case on "fm" in the section title here
\label{sec:lfmgap}
In this section, we present the details of the proposed recommendation approach,
{\it GAPfm}. We first introduce a smoothed version of GAP, for which
we can use to optimize a latent
factor model. Further, we analyze the complexity of the
learning algorithm, and propose an adaptive selection strategy that 
achieves constant complexity for {\it GAPfm}.
\subsection{Smoothed Graded Average Precision}
As shown in Eq.~(\ref{eqn:gap}), GAP depends on the rankings of the items in the
recommendation lists. However, the rankings of the items are not smooth
with respect to the predicted user-item relevance, and thus, GAP
results in a non-smooth function with respect to the latent factors of users and
items, i.e., $U$ and $V$.  Therefore, standard optimization
methods cannot be used to maximize the objective function as in
Eq.~(\ref{eqn:gap}). 
%
%In this work we exploit core ideas from the literature of learning to
In this work, we exploit core ideas from the literature on learning to
rank~\cite{Chapelle2010} and recent work that successfully used smoothed 
approximations of evaluation metrics for CF with implicit
feedback data~\cite{Shi2012,Shi2012b}.
% 4U2CHECK: Something seems to be missing here.
We approximate the rank-based terms in the GAP metric with smoothed functions with respect to the model parameters (i.e., the latent factors of users and items).
Specifically, the rank-based terms $1/R_{mi}$ and $\mathbb{I}(R_{mj}\leq
R_{mi})$ in Eq.~(\ref{eqn:gap}) are approximated by smoothed functions
with respect to the model parameters $U$ and $V$, as shown below:
\begin{align}
\label{eqn:approx1}
\mathbb{I}(R_{mj} \leq R_{mi})\approx
g(f_{mj}-f_{mi})
\end{align}
\vspace{-2pt}
\begin{align}
\label{eqn:approx2}
\frac{1}{R_{mi}}\approx
g(f_{mi})
\end{align}
\noindent where $g(x)$ is a logistic function, i.e., $g(x)=1/(1+e^{-x})$. The
basic assumption of the approximation in Eq.~(\ref{eqn:approx1}) is validated
in~\cite{Chapelle2010}, i.e., the condition of item $j$ being ranked higher
than item $i$ is more likely to be satisfied, if item $j$ has relatively higher relevance score than item $i$. 
However, the approximation in Eq.~(\ref{eqn:approx2}) proposed
% 4U2CHECK: changed rational -> rationale
in~\cite{Shi2012,Shi2012b} is heuristic, and the rationale of this approximation has not been well justified.
In this paper, we present our theoretical analysis of the approximation in
Eq.~(\ref{eqn:approx2}) and support its validity. The detail of the validation
is given in Appendix~\ref{app:approx}.

We attain a
smoothed version of $GAP_m$ by substituting the approximations introduced in
Eq.~(\ref{eqn:approx1}) and (\ref{eqn:approx2}) into Eq.~(\ref{eqn:gap}), as
shown below:
\begin{align}
\label{eqn:smoothgap}
GAP_m \approx & \sum_{i=1}^N I_{mi}g(f_{mi}) \sum_{j=1}^{N}I_{mj} 
\beta_{mij}g(f_{m(j-i)})
\end{align}
Note that for notation convenience, we make use of the substitution
$f_{m(j-i)}:=\left \langle U_m,V_j\right \rangle-\left \langle U_m,V_i\right
\rangle$.
% 4U2CHECK: I usually just say "drop" and not "drop out"
In Eq.~(\ref{eqn:smoothgap}), we drop the coefficient $1/Z_m$,
% 4U2CHECK: , thus, having  -> and, thus, has
which is independent of latent factors, and thus, has no influence on the
optimization of $GAP_m$. Taking
% 4U2CHECK: I think it should be "the GAP of all"
into account GAP of all $M$ users (i.e., the average GAP across all the users)
and the regularization for the latent factors, we obtain the objective function of {\it GAPfm} as below:
\begin{align}
\label{eqn:f}
F(U,V)=& \frac{1}{M}\sum_{m=1}^M \sum_{i=1}^N I_{mi}g(f_{mi})
\sum_{j=1}^{N}I_{mj} \beta_{mij}g(f_{m(j-i)}) \nonumber \\
&  -\frac{\lambda}{2}(\|U\|^2+\|V\|^2)
\end{align}
$\|U\|$ and $\|V\|$ are Frobenius norms of $U$ and $V$, and $\lambda$ is the
parameter that controls the magnitude of regularization. Note that the constant
coefficient $1/M$ is also dropped in the following, since it has no influence on the optimization of $F(U,V)$.

\subsection{Optimization}
\label{sec:opt}
Since the objective function in Eq.~(\ref{eqn:f}) is smooth over the model
parameters $U$ and $V$, we can optimize it using stochastic gradient ascent.
In each iteration, we optimize
$F(U_m,V)$ for user $m$ independently of all the other users. 
The gradients of $F(U_m,V)$ with respect to user $m$ and item $i$ can
be computed as follows:
{\small
\begin{align}
\label{eqn:fdu}
\frac{\partial F}{\partial U_m}=& \sum_{i=1}^{N}I_{mi} \big [
g'(f_{mi})\sum_{j=1}^N I_{mj}\beta_{mij}g(f_{m(j-i)})V_i  \nonumber \\
&+ g(f_{mi}) \sum_{j=1}^N
I_{mj}\beta_{mij}g'(f_{m(j-i)})(V_j-V_i) \big]
-\lambda U_m
\end{align}
\begin{align}
\label{eqn:fdv}
\frac{\partial F}{\partial V_i}= & I_{mi}\big [ g'(f_{mi})
\sum_{j=1}^{N}I_{mj} \beta_{mij}g(f_{m(j-i)}) \nonumber \\
&+ \sum_{j=1}^N I_{mj}
(\beta_{mji}g(f_{mj})-\beta_{mij}g(f_{mi}))g'(f_{m(j-i)})
\big]U_m \nonumber \\
& -\lambda V_i
\end{align}
}
The derivation of the gradient in Eq.~(\ref{eqn:fdu}) is rather straightforward.
However, the derivation of Eq.~(\ref{eqn:fdv}) is more sophisticated, since the
latent factors of different items are coupled. For this reason, we leave its
complete derivation in Appendix~\ref{app:fdv}.

The complexity of computing the gradient in
Eq.~(\ref{eqn:fdu}) is $O(DS_m^2+D)$, where $S_m$ denotes the number of items
rated/graded by user $m$. Taking into account all the $M$ users, the complexity
of computing gradients in
Eq.~(\ref{eqn:fdu}) in one iteration can be denoted as $O(D\overline{S}^2M+DM)$,
in which $\overline{S}$ is the average number of rated items across all the users.
Similarly, for a given user $m$, the complexity of computing the gradient in
Eq.~(\ref{eqn:fdv}) is $O(DS_m^2+DS_m)$, and the complexity in one iteration
over all the users is $O(D\overline{S}^2M+D\overline{S}M)$. Note that since we
have the conditions $|Y|=\overline{S}M$ and $|Y|>>M,\overline{S},D$, the overall
complexity of {\it GAPfm} in one iteration can be regarded as $|Y|$, which is
linear to the number of observed ratings in the given dataset.

In addition, as can be observed, for a fixed set of latent factors
$V$, updating the latent factors of user $m$ as in Eq.~(\ref{eqn:fdu}) can be
% 4U2CHECK: I would say "carried out" and not "conducted" here
conducted independently of updates of all the other users. Therefore, in each
iteration, the optimization of $U$ for individual users can be done in parallel.
As a result, the time complexity of updating $U$ in practice can even be a
constant, which is determined by the computing facilities (i.e., number of
processors), while not bounded by the scale of $|Y|$. In Section~\ref{sec:exp},
we will experimentally show the exploitation of parallel computing for
{\it GAPfm} and the resulting benefit.

However, as shown in Eq.~(\ref{eqn:fdv}), the latent factors of one item, e.g.,
% 4U2CHECK: is should be are here
item $i$, is coupled with the latent factors of some other items. Therefore, the
advantages of using parallel computing for updating $U$ cannot be
transferred to the update of $V$. Although the linear complexity with
respect to the scale of the given dataset $|Y|$ is already a crucial advantage
for the scalability of {\it GAPfm}, we shall still investigate possibilities to
further reduce the computational complexity, especially for the cases when the data is of
extremely large scale.

\subsection{Adaptive Selection}
\label{sec:adapt}
\label{sec:fast-learning}
%%LB we need an experiment where we show that this is better than simply selecting K random items.
% 4U2CHECK: I would say "is to adaptively select" instead of "is by adaptively selecting"
% 4U2CHECK: Does it make sense to mentioned related work on such techniques?
The key idea we propose to reduce the complexity of updating $V$ is to
adaptively select a fixed number ($K$) of items for each user in each
iteration and only update their latent factors. The criterion of adaptive
selection is to select the $K$ most ``misranked'' (or disordered) items for
each user in each iteration. We present the theoretical support for this
technique in
Appendix~\ref{app:adapt}. For example, if user $m$ has three rated items with
ratings 2, 4, 5 (i.e., the rank of the third item is 1), and the predicted relevance scores (i.e., $f_{mi}$) after an iteration are 0.3, 0.5, 0.1 (i.e.,
the rank of the third item is predicted to be 3), then the third item is the
most ``misranked'' one, which (assuming $K$ is set to 1) will be selected for
updating its latent factors in the next iteration. Essentially, the $K$ items are
adaptively selected representing the worst offending items with
respect to the loss function. The optimization of
the latent factors of these selected items yields the highest benefit in terms of optimizing GAP.
We summarize the adaptive selection strategy in
Algorithm~\ref{alg:adapt}. 

Note that with a fixed $K$, the complexity of updating $V$ in one iteration is
$O(DK^2M+DKM)$. As a result, using $C=KM$ (note that $C>>D,K$), the overall
complexity of {\it GAPfm} is in the magnitude of $O(C)$, which is a constant being independent of
the scale of $|Y|$. In Section~\ref{sec:exp}, we will experimentally validate
the complexity of {\it GAPfm} and the usefulness of the adaptive selection strategy.
Summarizing, the entire learning algorithm of {\it GAPfm} is illustrated in
Algorithm~\ref{alg:lfmgap}.

\begin{algorithm}[t]
\SetAlgoNoLine
\KwIn{Graded items by user $m$, i.e., $N_m=\{y_{mi}>0, i=1,2,\ldots,N\}$,
latent factors $U_m$ and $V$, and the number of selected items $K$.}
\KwOut{Selected item set for user $m$: $T_{m}$.}
		\If{$\sum_{i=1}^NI_{mi}\leq K$}
        {$T_m=N_m$\;
        break \;}
Compute $r$ as a vector representing the ranks of items in $N_m$ according to
the descending order of $y_{mi}$\;
Compute $\hat{r}$ as a vector representing the ranks of items in $N_m$ according
to the descending order of $\left \langle U_m,V_i\right \rangle$\;
$dist=abs(r-\hat{r})$; \% a vector of absolute error between $r$ and $\hat{r}$
Sort elements of $dist$ in descending order\;
Set $idx$ as the indexes of top $K$ elements in $dist$\;
$T_m=N_m[idx]$\;
\caption{Fast Learning: AdaptiveSelection}
\label{alg:adapt}
\end{algorithm}

% Algorithm
\begin{algorithm}[t]
% \SetAlgoNoLine
\SetAlgoVlined
\KwIn{Training set $Y$, the number of items
for adaptive selection $K$, regularization parameter $\lambda$, learning rate
$\gamma$, and the maximal number of iterations $itermax$.}
\KwOut{The learned latent factors $U$ and $V$.}
\For{$m=1,2,\ldots,M$
    }{
    \% Index graded items from each user\;
      $N_m=\{i|y_{mi}>0,i=1,2,\ldots,N\}$\; }
Initialize $U^{(0)}$ and $V^{(0)}$ with random values, and
$t=0$\; 
\Repeat{$t \geq itermax$}{
		\% Updating $U$ in parallel\;
        \For{$m=1,2,\ldots,M$
    }{
      $U_m^{(t+1)}=U_m^{(t)}+\gamma \frac{\partial F}{\partial U_m^{(t)}}$ 
      based on Eq.~(\ref{eqn:fdu})\; }
      \% Updating $V$\;
        \For{$m=1,2,\ldots,M$
    }{
    $T_m=AdaptiveSelection(N_m,U_m^{(t)},V^{(t)},K)$\;
    \For{$i\in T_m$}{
      $V_i^{(t+1)}=V_i^{(t)}+\gamma \frac{\partial F}{\partial V_i^{(t)}}$ based
      on Eq.~(\ref{eqn:fdv})\; }
      }
        $t=t+1$\;}
      $U=U^{(t)}$, $V=V^{(t)}$\;
\caption{{\it GAPfm}}
\label{alg:lfmgap}
\end{algorithm}

\subsection{Discussion}
In addition to the aforementioned learning algorithm of {\it GAPfm} and its
complexity analysis, we further discuss two characteristics of GAP
% 4U2CHECK: I changed: that could illustrate more insights -> that provide more insight into
that provide more insight into the usefulness of {\it GAPfm}.

First, since GAP can be regarded as a generalization of AP to
multi-grade data~\cite{Robertson2010}, {\it GAPfm} can be also seen as a
generalization of~\cite{Shi2012}, a CF approach that directly optimizes
AP in the implicit feedback domain. This can also be seen by looking at the smoothed
version of GAP in Eq.~(\ref{eqn:smoothgap}), for the case of $y_{max}=1$. For better
readability, we leave the detailed proof in Appendix~\ref{app:char}. This
characteristic indicates that although {\it GAPfm} is specifically designed for
the recommendation domains with graded relevance data, it can also be
utilized for the optimization of AP in the domains with implicit feedback data ,
as it becomes equivalent (for $y_{max} =1$) to the approach proposed
in~\cite{Shi2012}.

Second, since GAP is an approximation
to the area under the graded precision-recall curve as illustrated
in~\cite{Robertson2010}, {\it GAPfm} can also be extended to the
optimization of graded precision (GP) and graded recall (GR) at the
top-N part of the recommendation list, as shown in Appendix~\ref{app:char}. Note
that similar to {\it GAPfm}, we can first approximate GP@n and GR@n by smoothed functions of latent factors, and then learn the latent factor
models in the same fashion as approached in {\it GAPfm} (c.f.
Section~\ref{sec:opt} and \ref{sec:adapt}). This characteristic of
{\it GAPfm} is important
for real-world applications, since {\it GAPfm} can be simply adapted in
real-world systems and tuned to achieve high recall or precision at certain position
%ML changed: of users' recommendation lists. We also notice that our
%approach can also be applied for optimizing another recently
%proposed evaluation metric, expected reciprocal rank~\cite{Chapelle2009}, while in this
%paper, our focus remains in the optimization of GAP and its properties. In our
%experiments, we focus on the evaluation of our main contribution, {\it GAPfm}.
of users' recommendation lists. We also note that, beyond the scope of the current paper, our
approach could also be applied for optimizing another recently
proposed evaluation metric, expected reciprocal rank~\cite{Chapelle2009}.
%4YS: Please check
Here, we keep our focus on
the optimization of GAP, and on understanding the properties of {\it GAPfm} and evaluating its performance. 
%In our
%experiments, we focus on the evaluation of our main contribution, {\it GAPfm}.

\section{Experimental Evaluation}
\label{sec:exp}
In this section we present a series of experiments to evaluate the proposed
{\it GAPfm} algorithm. We first give a detailed description of the dataset and setup that
is used in the experiments. Then, we validate several properties of
{\it GAPfm}, as mentioned in Section~\ref{sec:lfmgap}. Finally, we
compare the recommendation performance between {\it GAPfm} and
several state-of-the-art alternative approaches.

We design the experiments in order to address the following
research questions:
1) Is {\it GAPfm} effective for optimizing GAP?
2) Is {\it GAPfm} scalable for large-scale use cases?
3) Does {\it GAPfm} outperform state-of-the-art CF approaches for top-N
recommendation?

\subsection{Experimental Setup}
\subsubsection{Dataset}
\label{sec:data}
The Netflix Prize dataset is one of the most used graded relevance dataset for
CF\footnote{http://en.wikipedia.org/wiki/Netflix\_Prize}. Two parts of the dataset are used, i.e., the training set and the probe set. The
training set consists of ca. 99M ratings (integers scaled from 1 to 5)
from ca. 480K users to 17.7K movies. The probe set contains ca. 1.4M ratings
disjoint from the training set. The training set is used for
building recommendation models, and the probe set is used for the evaluation. In
the experiments, we exclude the users with less than 50 ratings from the
training set. This choice is made to guarantee that users have sufficient number
of ratings so as to
facilitate our investigation on different cases in terms of the number of
ratings per user. This also allows us to analyze the complexity of {\it GAPfm}, as
discussed in Section~\ref{sec:fast-learning}. Note that this exclusion
criterion removes  one
third of users, without dramatically reducing the size of the dataset,
i.e., it only results in an reduction of 4\% of the number of
ratings. 
The statistics of the training set are summarized in
Table~\ref{tab:data}.

\begin{table}[t]% \scriptsize 
%\vspace{-2mm}
\small
\centering
\caption{\small Statistics of the training set in the experiments.}
\label{tab:data}
\begin{tabular}{c c c c}
\hline
\# users & \# movies & \# ratings & Sparseness \\
\hline
319275 & 17770 & 95085018 & 98.32\% \\
\hline
\end{tabular}
% \vspace{-2mm}
\end{table}

\subsubsection{Experimental Protocol}
\label{exp:protocol}
We randomly select a certain number of rated movies and their ratings for each
user in the training set to form a training data fold. For example, under the condition of
``Given 10'', we randomly select 10 rated movies for each user in
order to generate
a training data fold, from which the ratings are then used as the input data to
train the recommendation models. In the experiments we investigate a
variety of ``Given'' conditions, i.e., 10, 20, 30, and 50. Based on the learned
recommendation models, recommendation lists can be generated for each user, and the performance can be measured according to
the ground truth in the probe set. Note, that the probe set was originally
designed for the purpose of measuring the accuracy of rating prediction in the
Netflix Prize competition. However, it is not guaranteed that for any user the
performance of a recommendation list is measurable. For example,
it is infeasible to measure the performance of a ranked list 
if a user has only one rating in a probe set.
For this reason, in our experiments we
choose to measure the recommendation performance only based on the ground truth
of the users who have at least 5 rated movies in the probe set (ca. 14\% users
in the probe set).
Note that the choice of 5 is set to allow all the evaluation metrics (as introduced
later in this section) to achieve the highest possible value $1$ for the
task of top-5 recommendation. 

In addition to GAP, two additional evaluation metrics are used in the
experiments to measure the recommendation performance, i.e., NDCG and Precision.
As mentioned in Section~\ref{sec:intro}, NDCG is a well-known evaluation metric
for measuring the performance of ranked results with graded relevance. Precision
is a traditional evaluation measure that reflects the ratio of relevant items in
the ranked list given a truncated position. A relevance threshold of determining
relevant items is necessary when precision is applied to the graded relevance
scenarios. We set the relevance threshold to be 5 (the
highest rating value in the dataset), when measuring the precision of the
recommendation list. This choice is also supported by literature on
recommender systems evaluation~\cite{Cremonesi2010}. 

Since in recommender systems the user's satisfaction is dominated by only a few
items on the top of the recommendation list, our evaluation in the following
experiments focuses on the performance of the top-5 recommended items, i.e., GAP@5,
NDCG@5 and P@5 (averaged across all the test users) are used to measure the
recommendation performance. Note that in the evaluation we cannot treat  all the unrated items/movies as
irrelevant to a given user. A widely-used practical strategy,~\cite{Cremonesi2010,Koren2008,Shi2012}, is to first randomly select
1000 unrated items (which are assumed to be irrelevant to the user) in addition to the ground truth (i.e., rated items) for each user in
the probe set, and then evaluate the performance of the recommendation list that
consists of only the selected unrated items and the ground truth items. This
evaluation strategy is also adopted here.

We randomly select 1.5\% (a similar size to the probe set) of the data in the
training set to generate a validation set, which is used to determine parameters that are involved in
{\it GAPfm} and baseline approaches and also to investigate the properties of
{\it GAPfm} as presented in the next section. 

\paragraph{Parameter setting} We set the dimensionality of
latent factors to be 10 for both {\it GAPfm} and other baseline approaches based
on latent factor models. The remaining parameters are empirically tuned so as
to yield the best performance in the validation set, i.e., for {\it GAPfm} we set the regularization parameter $\lambda$=0.001 and the learning rate $\gamma$=$10^{-5}$.

\paragraph{Implementation}
We implement the proposed {\it GAPfm} in Matlab using its parallel computing
toolbox, and run the experiments on a single PC with Intel Core-i7 (8 cores)
2.3GHz CPU and 8G RAM memory. For the purpose of reproducibility, we make
our implementation code of {\it GAPfm} publicly available (link blinded for review).

\subsection{Validation: Effectiveness}
\begin{figure}[t]
\centerline{
	\includegraphics[width=7.5cm]{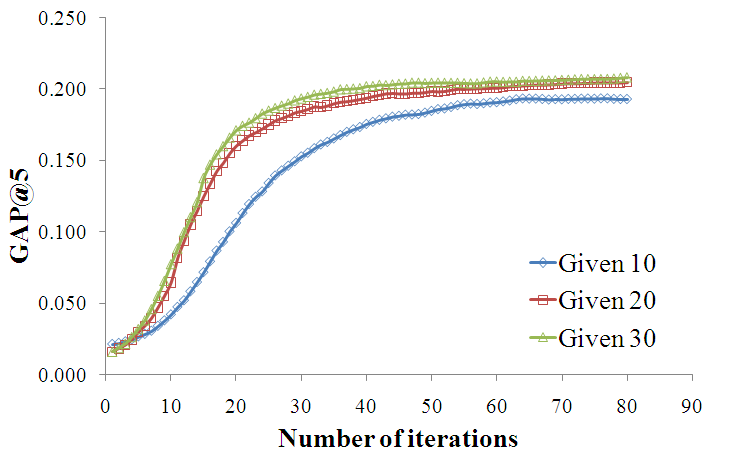}
}
% \vspace{-1.2em} % TODO: check this space hack
\caption{\small The Effectiveness of {\it GAPfm}.}
\label{fig:effect}
% \vspace{-3mm}
\end{figure}

In the first experiment we investigate the effectiveness of {\it GAPfm}, i.e.,
whether learning latent factors based on {\it GAPfm} contributes to the
improvement of GAP. We use the training set under the conditions ``Given 10'',
``Given 20'' and ``Given 30'', respectively, to train the latent factors, $U$
and $V$, which are then used to generate recommendation lists for individual users. The
performance of GAP is measured according the hold-out data in the validation set
along the iterations of the learning algorithm as described in Section~\ref{sec:lfmgap}.
The results are shown in Fig.~\ref{fig:effect}, which demonstrates that GAP
gradually improves along the iterations and attains convergence after a certain
number of iterations. For example, under the condition of ``Given 10'',
it converges after 60 iterations, while it converges with less iterations as more
data from the users is available for training. According to the observation from
this experiment, we can confirm a positive answer to our first research
question.

\subsection{Validation: Scalability}
\label{sec:scale}
We conduct three experiments to validate the scalability of
{\it GAPfm}. In the
first experiment we empirically investigate the benefits that we can draw from employing parallel computing for updating latent user factors in
{\it GAPfm}. We then validate the overall complexity of {\it GAPfm}  
as theoretically analyzed in
Section~\ref{sec:opt}. The last experiment is conducted to investigate the impact of the
proposed adaptive selection strategy, as discussed in Section~\ref{sec:adapt}.

\subsubsection{Parallel Updating of Latent User Factors}
\begin{figure}[t]
\centerline{
	\includegraphics[width=7cm]{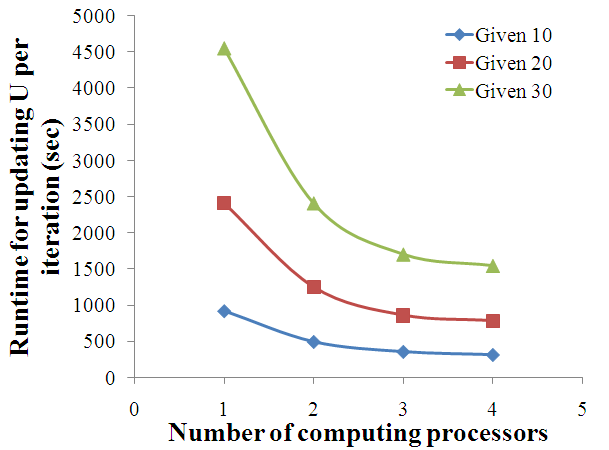}
}
% \vspace{-1.2em} % TODO: check this space hack
\caption{\small The runtime of updating $U$ in parallel in the
learning algorithm of {\it GAPfm}.}
\label{fig:par}
% \vspace{-3mm}
\end{figure}

As shown in Section~\ref{sec:opt}, the update of latent user factors $U$ in
{\it GAPfm} can be conducted in parallel. By utilizing multiple cores/processors
of the computing machine for the experiments, we empirically investigate the
average runtime for updating $U$ per iteration against the number of processors
employed. Note that updating $U$ in parallel has no influence on the quality of
the resulting latent factors, and thus, the performance of the recommendation lists
is not influenced. As shown in Fig.~\ref{fig:par}, for each of the ``Given''
conditions, the runtime of updating $U$ in the learning algorithm is
reduced remarkably
when increasing the number of processors for parallelizing the learning
process. Note that the reduction of runtime for updating $U$ is not exactly
inversely-proportionate to the number of processors, due to the system
process maintenance constraints.
%For example, although the computing machine used in our experiments has 8 cores,
%we do not observe remarkable runtime drop when using more than 4 cores. 
%%LB Why there is only 4 CPUS and not 8? 
However,
the experiments are sufficient to demonstrate that parallelization
can be used to speedup the optimization of the latent user factors in
{\it GAPfm}, given adequate computing facilities.

\subsubsection{Linear Complexity}
\begin{figure}
\centerline{
	\includegraphics[width=7.5cm]{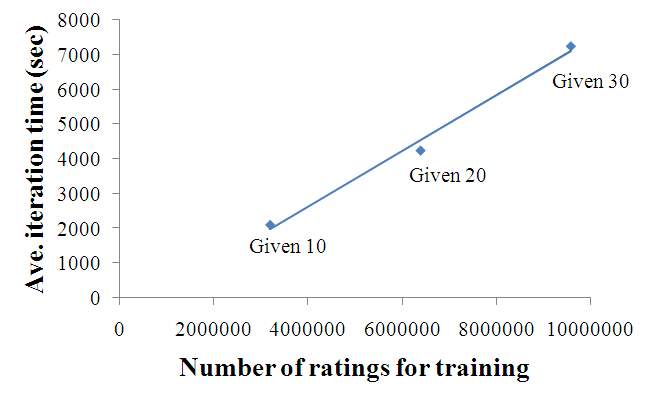}
}
% \vspace{-1.2em} % TODO: check this space hack
\caption{\small The relationship of the average iteration time against the size
of data for training {\it GAPfm}.}
\label{fig:linear}
% \vspace{-3mm}
\end{figure}

Since the size of the data used for training the latent factors in {\it GAPfm} varies
under different ``Given'' conditions, we can empirically measure the complexity
of {\it GAPfm} by observing the computational time consumed at each condition.
In Fig.~\ref{fig:linear}, the average iteration time is shown, which grows nearly
linearly as the number of ratings used for training increases. This result validates the
scalability of {\it GAPfm} under the basic setting, i.e., it scales
linearly to the amount of observed data in the training dataset.

\subsubsection{Impact of Adaptive Selection}
%%LB would be good to compare with the random strategy, I guess time should be better, GAP worse... but where do they intersect?
In order to investigate the impact of the adaptive selection strategy for
{\it GAPfm}, we design an experiment under the condition of ``Given 50" to measure the computational
cost for the cases where different numbers of items, i.e., different values for
$K$ (varying from 5 to 50), are adaptively selected during iterations for each
user for training {\it GAPfm}. Meanwhile, we measure the performance of GAP based on the hold-out
%4U2CHECK I usually call it "held-out data" maybe check this usage...
data in the validation set, corresponding to each case of $K$. 

\begin{figure}
\centerline{
	\includegraphics[width=7.5cm]{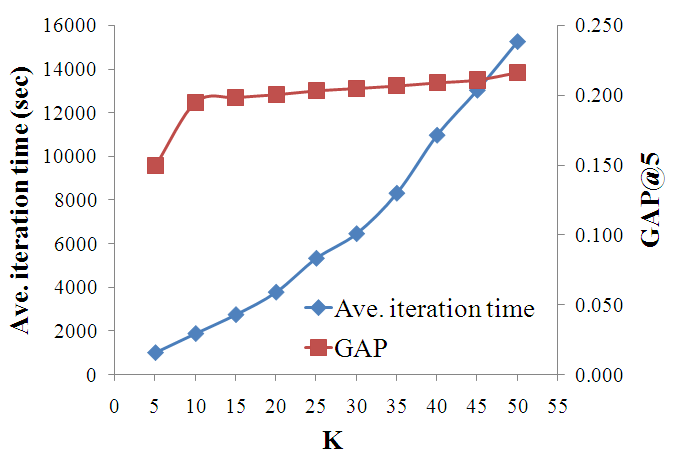}
}
\caption{\small The impact of value $K$ on both the computational cost and the
performance of GAP.}
\label{fig:adapt}
% \vspace{-3mm}
\end{figure}

The results are
shown in Fig.~\ref{fig:adapt}. Note that increasing $K$ is equivalent to
increasing the size of the data for training {\it GAPfm}. Thus, as
shown in the
%4U2CHECK in the next sentence I think you mean "the previous experiment" or "previous experiments"
previous experiment, the average iteration time increases almost linearly to the
growth of $K$. However, we do observe that {\it GAPfm} already achieves
relatively high GAP even with a small value of $K$, compared to the GAP
achieved when the latent factors of all the items (i.e., $K=50$) are updated.
For instance, when only 20 items are adaptively selected for updating their
latent factors in each iteration of the learning algorithm, i.e., $K=20$,
nearly 75\% of the runtime can be saved, compared to the case of updating the
latent factors of all the items. Meanwhile, the drop of GAP is only around 5\%,
i.e., the GAP is 0.206 when $K=20$, and 0.216 when $K=50$. For
%4U2CHECK applying needs a direct object, maybe say "When moving to..."
large-scale datasets where computational cost is crucial, the small performance hit introduced by
the adaptive selection process pays off due to the large gain attained
in terms of computational
cost. As analyzed in Section~\ref{sec:adapt}, we can practically maintain a
constant complexity of {\it GAPfm} by using adaptive selection with a fixed $K$.
In next section, we will further demonstrate the performance of {\it GAPfm} with
adaptive selection in the case of $K=20$. Finally, we also conducted an
experiment for validating the utility of adaptive selection, compared to
random selection, i.e., in the case of $K=20$ we randomly select 20 rated
items for updating their latent factors in each iteration of {\it
  GAPfm}. Random selection for $K=20$ produced a  value of GAP = 0.166, which is 19\% lower than computed using adaptive
selection. In total, the experimental results confirm the value of the proposed
adaptive selection for {\it GAPfm} in terms of both recommendation performance
and the computational cost.

Summarizing, the observations from the experiments presented above allow us to
give a positive answer to our second research question, i.e., {\it GAPfm} can be
highly scalable (with constant complexity independent of the scale of the given
dataset) and used for large-scale datasets.

\begin{table*}[th]% \scriptsize 
%\vspace{-2mm}
\small
\centering
\caption{\small Performance comparison of {\it GAPfm} and the baseline approaches on
the Netflix dataset.}
\label{tab:perfnf}
\begin{tabular}{c|c c c|c c c|c c c}
\hline
 &  & Given 10 &  &  & Given 20 &  &  & Given 30 & \\
 & P@5 & NDCG@5 & GAP@5 & P@5 & NDCG@5 & GAP@5 & P@5 & NDCG@5 & GAP@5\\
 \hline
PopRec & 0.023 & 0.080 & 0.165 & 0.024 & 0.084 & 0.171 & 0.026 & 0.085 & 0.176\\
SVD++ & 0.023 & 0.065 & 0.133 & 0.027 & 0.071 & 0.139 & 0.032 & 0.087 &
0.168\\
CofiRank & 0.027 & 0.093 & 0.181 & 0.027 & 0.088 & 0.182 & 0.025 & 0.088 & 0.186\\
{\it GAPfm} & {\bf 0.040} & {\bf 0.109} & {\bf 0.195} & {\bf 0.042} &
{\bf 0.111} & {\bf 0.207} & {\bf 0.042} & {\bf 0.114} &
{\bf 0.213}\\
\hline
\end{tabular}
% \vspace{-2mm}
\end{table*}

\subsection{Performance on Top-N Recommendation}
\label{sec:perf}
We compare the performance of {\it GAPfm} with three baseline approaches. Each of the baseline approaches is listed and
briefly introduced below:
\begin{packed_item}
  \item \textbf{PopRec} is a naive and non-personalized baseline that recommends
  movies in terms of their popularity, i.e., the number of ratings
  from all the users. Note that another naive baseline based on the average
  rating of movies was also tested, but it achieved rather low performance
  compared to other baselines. For this reason, we excluded it from our
  experimental results reported in this paper. Although being a naive baseline,
  PopRec is shown in literature to have competitive performance for the top-N
  recommendation task on the Netflix dataset~\cite{Cremonesi2010}.
  \item \textbf{SVD++} is a state-of-the-art CF approach, which is shown
  to be superior for the rating prediction task as witnessed in the Netflix
  Prize contest~\cite{Koren2008}. We use the implementation of SVD++ available in
  GraphLab\footnote{http://graphlab.org/toolkits/collaborative-filtering/}~\cite{Low2012}.
  Note that the dimensionality of latent factors used in SVD++ is set to 10, the
  same as the setting for the proposed {\it GAPfm}. Other parameters, such as the regularization parameter, are tuned according to the observation from the validation set. 
  \item \textbf{CofiRank} is a state-of-the-art CF approach that
  specifically optimizes for the ranking performance of recommendation
  results~\cite{Weimer2007}. In addition, to the best of our knowledge, CofiRank
  is also the only well-established CF approach that directly optimizes NDCG measure
  for graded relevance datasets. Note that the implementation of
  CofiRank is based on the publicly available software package from the
  authors\footnote{http://www.cofirank.org/}. The dimensionality of latent
  factors is also set to 10, and the remaining parameters are tuned on the validation
  set.
\end{packed_item}
\vspace{-2.5mm}
% Note that in this experiment we did not include the collaborative ranking
% approaches as baselines, since those approaches reported in literature are
% evaluated by the performance of ranking rated items (not exactly the top-N
% recommendation), as discussed in Section~\ref{sec:rw}. For instance, we have
% tested CofiRank~\cite{Weimer2007} as another baseline, which is well performed
% for the ranking measures based on the rated items in the probe set, while performing even
% slightly worse than the other two baselines reported in this experiment. For
% this reason, we leave the comparison with collaborative ranking approaches to
% another experiment, as shown in next subsection.

In Table~\ref{tab:perfnf}, we present the performance of {\it GAPfm} and the
baseline approaches for ``Given'' 10 to 30 items per user in the training set. Under all three
conditions, {\it GAPfm} largely outperforms all the baselines, i.e., over 30\% in P@5,
15\% in NDCG@5 and 10\% in GAP@5. All the improvements are statistically
significant, according to Wilcoxon signed rank significance test with p<0.01.
The results indicate that the proposed {\it GAPfm} is highly competitive for the top-N
recommendation task. We also demonstrate that the optimization of GAP leads to improvements in terms of precision and
NDCG. We also notice that SVD++ is only slightly better than PopRec in P@5
(which was also observed in related work of ~\cite{Cremonesi2010}), but worse
than PopRec in both NDCG@5 and GAP@5. This result again indicates that
optimizing rating predictions do not necessarily lead to good performance for
top-N recommendations.

In Table~\ref{tab:perfadapt}, we show  the performance of {\it GAPfm}
with adaptive selection for ``Given 50'' items per user in the training set, which simulates the case of relatively large user
profiles. As indicated in Section~\ref{sec:scale}, we adopt $K=20$ for the
adaptive selection in {\it GAPfm}. 
{\it GAPfm} still achieves a large improvement over all of the baselines across all
the metrics, and also slightly outperforms {\it GAPfm} with adaptive selection, i.e., ca.
6\% in NDCG@5 and ca. 8\% in GAP@5. However, we observe that {\it GAPfm} with adaptive
selection still improves over PopRec, SVD++ and CofiRank to a significant
extent. Moreover, as mentioned before, under the condition of ``Given 50", training
{\it GAPfm} with adaptive selection at $K=20$ saves around 75\% computation time.
%4U2CHECK: I tweaked this next sentence.
Therefore, the drop in recommendation performance can be considered to be acceptable for
real-world applications.

\begin{table}[t]% \scriptsize
%\vspace{-2mm}
\small
\centering
\caption{\small Performance of {\it GAPfm} with adaptive selection ($K=20$) on the
Netflix dataset under the condition ``Given 50''}
\label{tab:perfadapt}
\begin{tabular}{c c c c}
\hline
 & P@5 & NDCG@5 & GAP@5\\
 \hline
PopRec & 0.025 & 0.085 & 0.172\\
SVD++ & 0.031 & 0.080 & 0.150\\
CofiRank & 0.023 & 0.084 & 0.188\\
{\it GAPfm} & 0.044 & 0.122 & 0.219\\
{\it GAPfm}+Adaptive Selection & 0.044 & 0.115 & 0.201\\
\hline
\end{tabular}
% \vspace{-4mm}
\end{table}

\begin{table*}[ht!]% \scriptsize 
%\vspace{-2mm}
\small
\centering
\caption{\small NDCG performance of {\it GAPfm} compared to WLT 
on the MovieLens 100K dataset}
\label{tab:perfrated}
\begin{tabular}{c|c c c|c c c|c c c|c c c}
\hline
 &  & Given 10 &  &  & Given 20 &  &  & Given 30 &  &  & Given 40 & \\
 & @1 & @3 & @5 & @1 & @3 & @5 & @1 & @3 & @5 & @1 & @3 & @5\\
 \hline
WLT & 0.710 & 0.683 & 0.680 & 0.703 & 0.695 & 0.692 & 0.714 & 0.712 & 0.710 &
0.741 & 0.719 & 0.715\\
 {\it GAPfm} & 0.709 & \textbf{0.692} & \textbf{0.683} & \textbf{0.717} & 0.691 &
 \textbf{0.695} & \textbf{0.722} & 0.709 & 0.708 & 0.736 & 0.712 & 0.704\\
\hline
\end{tabular}
% \vspace{-2mm}
\end{table*}

\subsection{Performance on Ranking Graded Items}
\label{sec:perfrank}
The last experiment is conducted to examine the performance of {\it GAPfm} on the task of ranking a given list of rated/graded items.
We compare  {\it GAPfm} with other collaborative ranking approaches
proposed in the literature. Note that our focus on this paper is on top-N
recommendation,  which differs essentially from the ranking of rated items.
In this setting we do not sample unrated items but only focus on the correct ranking of the rated items.
%4U2CHECK: I think that the reader needs be reminded of what the essential difference is (see my comments before on being very clear also what the definitions of "collaborative ranking" is)
However, this experiment only serves to verify that {\it GAPfm} would
be still competitive even in the case of evaluating the ranking of rated items.
For this reason, we extend our experiment on a different dataset, i.e., the
MovieLens 100K
dataset\footnote{http://www.grouplens.org/node/73}~\cite{Herlocker1999}, and compare the performance of {\it GAPfm} with a state-of-the-art approach, Win-Loss-Tie (WLT) feature-based collaborative
ranking~\cite{Volkovs2012}, which is, to our knowledge, the latest contribution
to collaborative ranking. We follow exactly the same experimental protocol as
used in the work of~\cite{Volkovs2012}, to allow us to make a
straightforward comparison with the results reported in their work. 
The results are shown in Table~\ref{tab:perfrated}. We observe that {\it GAPfm} achieves competitive performance for
ranking rated items, across all the conditions and NDCG at all the truncation
levels. These results indicate that the optimization of GAP for top-N
recommendation would naturally lead to improvements in terms ranking graded items.

In summary, according to our observations from Section~\ref{sec:perf}
and~\ref{sec:perfrank}, we can conclude a positive answer to our last research
question.


\section{Conclusions and future work}
\label{sec:conclude}
We have presented {\it GAPfm}, a new CF approach for top-N recommendation,
%4U2CHECK by learning a latent factor model that directly optimizes GAP. We propose an
by learning a latent factor model that directly optimizes GAP. We propose an
%4U2CHECK adaptive selection strategy for {\it GAPfm} so that it could attain a constant
adaptive selection strategy for {\it GAPfm} so that it could attain a constant
computational complexity, which guarantees its usefulness for large scale use
scenarios. 
%4U2CHECK: stick with scenarios?
Our experiments also empirically validate the scalability of {\it GAPfm}. {\it GAPfm}
is demonstrated to substantially outperform the baseline approaches for the top-N
%4U2CHECK: I suggest changing largely -> substantially here
recommendation task, while also being competitive for the performance of ranking
graded items, compared to the state of the art.

There are a few directions for future work. First, inspired by statistical analysis of evaluation metrics~\cite{Wang2010b}, we would like to
analyze the relations and differences between learning methods that optimize different evaluation metrics. 
Second, we are also interested in developing distributed version of the
proposed {\it GAPfm}, by taking into account recent contributions of distributed
latent factor models~\cite{Gemulla2011}. Third, considering the multi-facet
relevance judgments in recommender systems, such as accuracy, diversity,
serendipity, we would also like to investigate the possibilities of optimizing
%4U2CHECK: what is cohesive? 
top-N recommendation with multiple cohesive or competing
objectives~\cite{Agarwal2012a,Jambor2010,Svore2011}.

\bibliographystyle{abbrv}
{
\scriptsize{\bibliography{MyRefs_all_sigir2013tid}}}

  % sigproc.bib is the name of the
%Bibliography in this case
% ACM needs 'a single self-contained file'!
%\balancecolumns
% \vspace{-4mm}
%\newpage
\appendix
%Appendix A
\section{Validation of Eq.~(7)}
\label{app:approx}
{\scriptsize
Note that we neglect the user index $m$ in the following, due to its irrelevance
to this validation. Given relevance predictions on all the items, i.e., $f_i$,
$i=1,2,\ldots,N$, the definition of the rank of item $i$, i.e., $R_i$, can be
formulated as below:
\begin{align*}
R_i=1+\sum_{j=1}^{N}\mathbb{I}(f_{i} < f_{j})
\end{align*}
Suppose that $f$ is continuous and bounded within $[a,b]$. We can approximate
$R_i$ by a smooth function of $f$ as below:
\begin{align*}
R_i \approx
1+\sum_{j=1}^{N}\frac{e^{f_j-f_i}}{e^{b}}=1+\frac{e^{-f_i}}{e^{b}}\sum_{j=1}^{N}e^{f_j}
\end{align*}
In the case $N$ is large (which is common for most recommendation domains), we
could further approximate the summation by integration over all the possible
values of $f$. Under this approximation, we obtain:
\begin{align*}
R_i &\approx 1+\frac{e^{-f_i}}{e^{b}}\int_{a}^{b}e^{f} \textup{d} f
= 1+\frac{e^{-f_i}}{e^{b}}(e^b-e^a) \\
&= 1+(1-e^{-(b-a)})e^{-f_i}
\end{align*}
Note that it is reasonable to assume $b-a>>0$, since theoretically $b$ could be
positive infinite and $a$ could be negative infinite (if latent factors are
drawn from real values). Therefore, we can finally obtain:
\begin{align*}
R_i \approx 1+e^{-f_i} 
\end{align*}
which validates the approximation in Eq.~(\ref{eqn:approx2}), i.e.,
$1/R_i=g(f_i)$.
}

\section{Derivation of Eq.~(11)}
\label{app:fdv}
{\scriptsize
Note that we drop the derivative of the
regularization term, i.e., $-\lambda V_i$ in the following. 
\begin{align*}
\frac{\partial F}{\partial V_i}
&= I_{mi}g'(f_{mi})U_m \sum_{j=1}^NI_{mj}\beta_{mij}g(f_{m(j-i)}) \\
&+ I_{mi}g(f_{mi})\sum_{j=1,j\neq i}^N I_{mj}\beta_{mij}g'(f_{m(j-i)})(-U_m) \\
&+\sum_{k=1,k\neq i}^N I_{mk}g(f_{mk})I_{mi}\beta_{mki}g'(f_{m(i-k)})U_m
\end{align*}
Replacing $k$ with $j$ in the last summation term, and applying the
property $g'(-x)=g'(x)$ so as to combine the last two summation terms, we
obtain:
\begin{align*}
\frac{\partial F}{\partial V_i}
&= I_{mi}g'(f_{mi})U_m \sum_{j=1}^NI_{mj}\beta_{mij}g(f_{m(j-i)}) \\
&+ U_m\sum_{j=1,j\neq i}^N I_{mi}I_{mj}\big ( 
\beta_{mji}g(f_{mj})- \beta_{mij}g(f_{mi}) \big ) g'(f_{m(j-i)})
\end{align*}
Note that $\beta_{mji}g(f_{mj})- \beta_{mij}g(f_{mi})=0$ if $i=j$. We can add
this condition into above and take out $I_{mi}$ and $U_m$ to attain
Eq.~(\ref{eqn:fdv}).
}

\section{Adaptive Selection}
\label{app:adapt}
{\scriptsize
We present our justification of the criterion for adaptive selection proposed in
Section~\ref{sec:adapt}. Note that in the following, we drop the user index $m$,
due to its irrelevance to this validation. By the definition of the indicator
function $\mathbb{I}$, we have $\mathbb{I}(y_i<y_j)+\mathbb{I}(y_j \leq y_i)=1$.
Therefore, we can rewrite Eq.~(\ref{eqn:subs}) as:
\begin{align*}
\beta_{ij}:&=
\big (1-\mathbb{I}(y_{j} \leq y_{i})\big ) \sum_{l=1}^{y_{i}}\delta_l  +
\mathbb{I}(y_{j} \leq y_{i}) \sum_{l=1}^{y_{j}}\delta_l \\
&= \sum_{l=1}^{y_{i}}\delta_l+
\big (\sum_{l=1}^{y_{j}}\delta_l-\sum_{l=1}^{y_{i}}\delta_l
\big )\mathbb{I}(y_{j} \leq y_{i})
\end{align*}
Substituting above equation into Eq.~(\ref{eqn:gap}), we obtain:
\begin{align*}
GAP=& \sum_{i=1}^N \frac{I_i \sum_{l=1}^{y_{i}}\delta_l}{R_i} \sum_{j=1}^N I_j
\mathbb{I}(R_{j} \leq R_{i}) \\
&+ \sum_{i=1}^N \frac{I_i}{R_i}\sum_{j=1}^N I_j \mathbb{I}(R_{j} \leq
R_{i}) \mathbb{I}(y_{j} \leq y_{i}) \big (\sum_{l=1}^{y_{j}}\delta_l-
\sum_{l=1}^{y_{i}}\delta_l \big)
\end{align*}
Note that the first summation is non-negative. Also note that when $R_{j} \leq
R_{i}$ and $y_{j} \leq y_{i}$ (i.e., two items are misranked), we have $\sum_{l=1}^{y_{j}}\delta_l-
\sum_{l=1}^{y_{i}}\delta_l \leq 0$. Therefore, the second summation is composed
by non-positive terms, i.e., only leading to a loss of GAP. In addition,
the loss is proportionate to $\sum_{l=1}^{y_{j}}\delta_l-
\sum_{l=1}^{y_{i}}\delta_l$, the difference of grades of misranked items.
Thus, it is obvious that the largest loss would be casued by the most misranked
item, justifying our criterion for adaptive selection, i.e., updating most
misranked items would lead to largest upgrade of GAP. }

\section{Characteristics of {\it GAPfm}}
\label{app:char}
{\scriptsize
\textbf{Generalization.} In the scenario with implicit feedback data (binary
relevance data), we have $y_{max}=1$. Thus, in the case of $I_{mi}>0$ and
$I_{mj}>0$, it implies that we have $y_{mi}=y_{mj}=y_{max}=1$. Substituting this
condition into Eq.~(\ref{eqn:subs}) and taking into account of the definition
of in Eq.~(\ref{eqn:gradeprop2}), we obtain $\beta_{mij}=1$. Furthermore, the
smoothed GAP as in Eq.~(\ref{eqn:smoothgap}) returns to:
\begin{align*}
GAP_m=\sum_{i=1}^N I_{mi}g(f_{mi}) \sum_{j=1}^N I_{mj}g(f_{m(j-i)})
\end{align*}
which is equivalent to the smoothed AP (excluding the context variable and the
constant coefficient) as proposed in the work of~\cite{Shi2012}. Therefore, the
proposed {\it GAPfm} is a generalization of the approach optimizing AP in the
implicit feedback domains.

\textbf{Specialization.} Graded precision and graded recall are two
specialized metrics that can be decomposed from GAP. As defined
in~\cite{Robertson2010}, we can rewrite GP@n and GR@n for the top-n
recommendation list of user $m$ as follows:
\begin{align*}
GP_m@n=\frac{1}{n}\sum_{i=1}^N I_{mi} \mathbb{I}(R_{mi} \leq n)
\xi_{mi}^{(n)} \\
GR_m@n=\frac{1}{Z_m}\sum_{i=1}^N I_{mi} \sum_{l=1}^{y_{mi}}
\delta_l \mathbb{I}(R_{mi} \leq n)
\end{align*}
where,
\begin{align*}
\xi_{mi}^{(n)}:=\frac{\mathbb{I}(y_{mi}<c_{mn})\sum_{l=1}^{y_{mi}}\delta_l+\mathbb{I}(c_{mn}\leq
y_{mi})\sum_{l=1}^{c_{mn}}\delta_l} {\sum_{j=1}^{c_{mn}}\delta_l}
\end{align*} 
$c_{mn}$ denotes the grade of the $n$th item in the user $m$'s ranked list,
according the descending order of her rated items. We may attain smoothed
versions of GP@n and GR@n by approximating $\mathbb{I}(R_{mi} \leq n)$ with
smoothed functions of latent factors $U$ and $V$, such as 
\begin{align*}
\mathbb{I}(R_{mi} \leq n)\approx g(\left \langle U_m,V_i\right \rangle-c_{mn})
\end{align*} 
Then, similar optimization as used in {\it GAPfm} can be performed to learned
latent factors that optimize GP@n or GR@n. }
%\balancecolumns % GM June 2007
% That's all folks!
\end{document}